\documentclass[10pt,a4paper,english,twocolumn]{IEEEtran}
\usepackage[T1]{fontenc}
\usepackage{color}
\usepackage{babel}
\usepackage{verbatim}
\usepackage{textcomp}
\usepackage{amsmath}
\usepackage{amstext}
\usepackage{graphicx}

\makeatletter

\pdfpageheight\paperheight
\pdfpagewidth\paperwidth

\usepackage{epstopdf}
\usepackage{cite}
\usepackage{citesort}
\usepackage{balance}

\@ifundefined{showcaptionsetup}{}{
 \PassOptionsToPackage{caption=false}{subfig}}
\usepackage{subfig}
\makeatother

\begin{document}

\title{On the Fundamental Characteristics of \\
Ultra-Dense Small Cell Networks
}

\author{\noindent {\normalsize{}Ming Ding}$^{*}$\thanks{$^{*}$Ming Ding is with Data61, CSIRO, Australia (e-mail: Ming.Ding@data61.csiro.au). }{\normalsize{},
}\textit{\normalsize{}Senior Member, IEEE}{\normalsize{}, David L$\acute{\textrm{o}}$pez-P$\acute{\textrm{e}}$rez}$^{\dagger}$\thanks{$^{\dagger}$David L$\acute{\textrm{o}}$pez-P$\acute{\textrm{e}}$rez
is with Nokia Bell Labs, Ireland (email: david.lopez-perez@nokia.com). }{\normalsize{}, }\textit{\normalsize{}Senior Member, IEEE}, {\normalsize{}}\\
{\normalsize{}Holger Claussen}$^{\ddagger}$\thanks{$^{\ddagger}$Holger Claussen is leader of the Small Cells Research
at Nokia Bell Labs, Ireland (email: holger.claussen@nokia.com). }{\normalsize{}, }\textit{\normalsize{}Senior Member, IEEE}{\normalsize{},
Mohamed Ali Kaafar$^{\mathsection}$}\thanks{$^{\mathsection}$Mohamed Ali Kaafar (Dali Kaafar) is the Group Leader
of Networks Group at Data61, CSIRO, Australia (e-mail: Dali.Kaafar@data61.csiro.au). }{\normalsize{}, }\textit{\normalsize{}Senior Member, IEEE}
}
\maketitle
\begin{abstract}
In order to cope with the forecasted 1000$\times$ increase in wireless capacity demands by 2030,
network operators will aggressively densify their network infrastructure to reuse the spectrum as much as possible.
However, it is important to realise that these new ultra-dense small cell networks are \emph{fundamentally different} from the traditional macrocell or sparse small cell networks,
and thus ultra-dense networks (UDNs) cannot be deployed and operated in the same way as in the last 25 years.
In this paper,
we systematically investigate and visualise the performance impacts of
several fundamental characteristics of UDNs,
that mobile operators and vendors should consider when deploying UDNs.
Moreover,
we also provide new deployment and management guidelines to address the main challenges brought by UDNs in the future.\footnote{To appear in IEEE Network Magazine. 1536-1276 copyright 2015 IEEE. Personal use is permitted, but republication/redistribution requires IEEE permission. Please find the final version in IEEE from the link: http://ieeexplore.ieee.org/document/xxxxxxx/. Digital Object Identifier: 10.1109/TNM.2017.xxxxxxx}
\end{abstract}

{}

\section{Introduction\label{sec:Introduction}}

Based on recent market predictions,
mobile data traffic will keep growing at an astonishing compound annual growth rate of 47\,\% towards 2030,
beyond the so-called 1000$\times$ wireless capacity demand~\cite{Report_CISCO}.
In response to this traffic explosion,
a number of technical studies have shown that a 1000$\times$ capacity gain can only be expected from solutions that involve the deployment of a massive number of cells,
either to \emph{i)} reuse the current spectrum (sub-6\,GHz solutions),
or \emph{ii)} gain access to new frequency bands (centimetre-wave and millimetre-wave solutions),
and the mobile industry has started to walk down this path~\cite{Tutor_smallcell,Ge2016UDNs}.

Throughout the second decade of 2000 up to now,
network densification has energised the 3rd Generation Partnership Project (3GPP) 4th-generation (4G) Long Term Evolution (LTE) networks,
with the full integration of small cell technology in their specifications,
and the commercial deployment of small cell products~\cite{Report_Mobile_Experts}.
Indoor small cells have conquered the residential market as a coverage solution.
Indeed, a total of over 14 million small cells were deployed worldwide by Feb. 2016,
of which over 12 million were residential.
The non-residential small cells business has also recently taken off,
both as a coverage and a capacity solution,
with enterprise and outdoor small cell shipments expected to grow by 270\,\% and 150\,\%, respectively, in 2017~\cite{Report_Mobile_Experts}.

\subsection{Orthogonal Deployments of UDNs in Sub-6\,GHz Spectrum\label{subsec:Role-of-UDNs}}

The current 5th-generation (5G) networks standardisation efforts and its stringent coverage and capacity requirements,
set to provide an unlimited user experience,
also suggest that small cells will remain the main driving force of cellular technology in the years to come,
with the overall small cell market expected to generate revenues of USD\,6.7 billion by 2020~\cite{Report_Mobile_Experts}.
In more detail,
\begin{itemize}
  \item
In the first deployment phase of 5G,
the orthogonal deployment\footnote{
     The orthogonal deployment means that small cells and macrocells are operating on a different frequency spectrum,
     i.e., 3GPP Small Cell Scenario \#2a~\cite{Tutor_smallcell}.}
of ultra-dense (UD) small cell networks (SCNs), a.k.a. ultra-dense networks (UDNs), at sub-6\,GHz frequencies is envisaged
as the workhorse for capacity enhancement for most operators,
mostly due to the large gains attainable via spatial reuse and the simplified interference and mobility management,
resulting from its low interaction with the macrocell network~\cite{Yunas2015UDNdep}.
This sub-6\,GHz small cell deployments are expected to achieve peak data rates of 1\,Gbps,
and a ubiquitous coverage with more than 100\,Mbps wherever needed.
  \item
In a posterior second phase of 5G,
the deployment of UDNs at centimetre-wave, millimetre-wave or even Tera-Hz~\cite{Barros2017THz} frequencies is expected to provide larger peak data rates of 10\,Gbps in megacities,
where fibre, digital subscriber line (DSL) and wireless based backhaul will be a widely available commodity~\cite{Ge2016UDNs}.
\end{itemize}

\subsection{Our Contributions and Paper Structure\label{subsec:Contribution}}

In this article,
we focus on the spatial reuse gain of UDNs  at sub-6GHz frequencies,
as their role outs are expected to be earlier than millimetre-wave ones\footnote{
We leave millimetre-wave communications analyses for further study as the topic stands on its own,
and requires a completely new analysis~\cite{Rangan2014mmWaveSurvey}.
More specifically,
millimetre-wave communications present several new features that should be treated differently from sub-6GHz ones,
such as short-range coverage, low inter-cell interference, blockage effects, and high Doppler shift, among others~\cite{Rangan2014mmWaveSurvey}.}.


Compared with the conventional sparse/dense cellular networks,
it is obvious that UDNs directly lead to
\begin{itemize}
\item {a much shorter distance between a base station (BS) and its served user equipment (UE)}, and
\item {a much smaller ratio of BSs to UEs. }
\end{itemize}
Following those two direct consequences,
in this article,
we present five characteristics that make UDNs fundamentally different from current sparse and dense networks,
and draw attention to those deployment aspects that operators will need to reconsider when deploying these UDNs.

The rest of this article is structured as follows:
In Section~\ref{sec:System-Model},
the most typical UDN use case together with a system model for its network performance study are described.
In Section~\ref{sec:Study-Methodology},
the network performance visualisation technique used in this paper is introduced.
In Sections~\ref{sec:two-fundamental-aspects-related-to-distance}$\sim$\ref{sec:one-fundamental-aspect-related-to-bursty-traffic},
the five fundamental differences that UDNs bring about are presented,
and their performance impacts are \emph{visualised}.
Finally, in Section~\ref{sec:Conclusion},
the conclusions are drawn.

\section{Use Case and System Model \label{sec:System-Model}}



\begin{figure}
\center
\includegraphics[width=8cm]{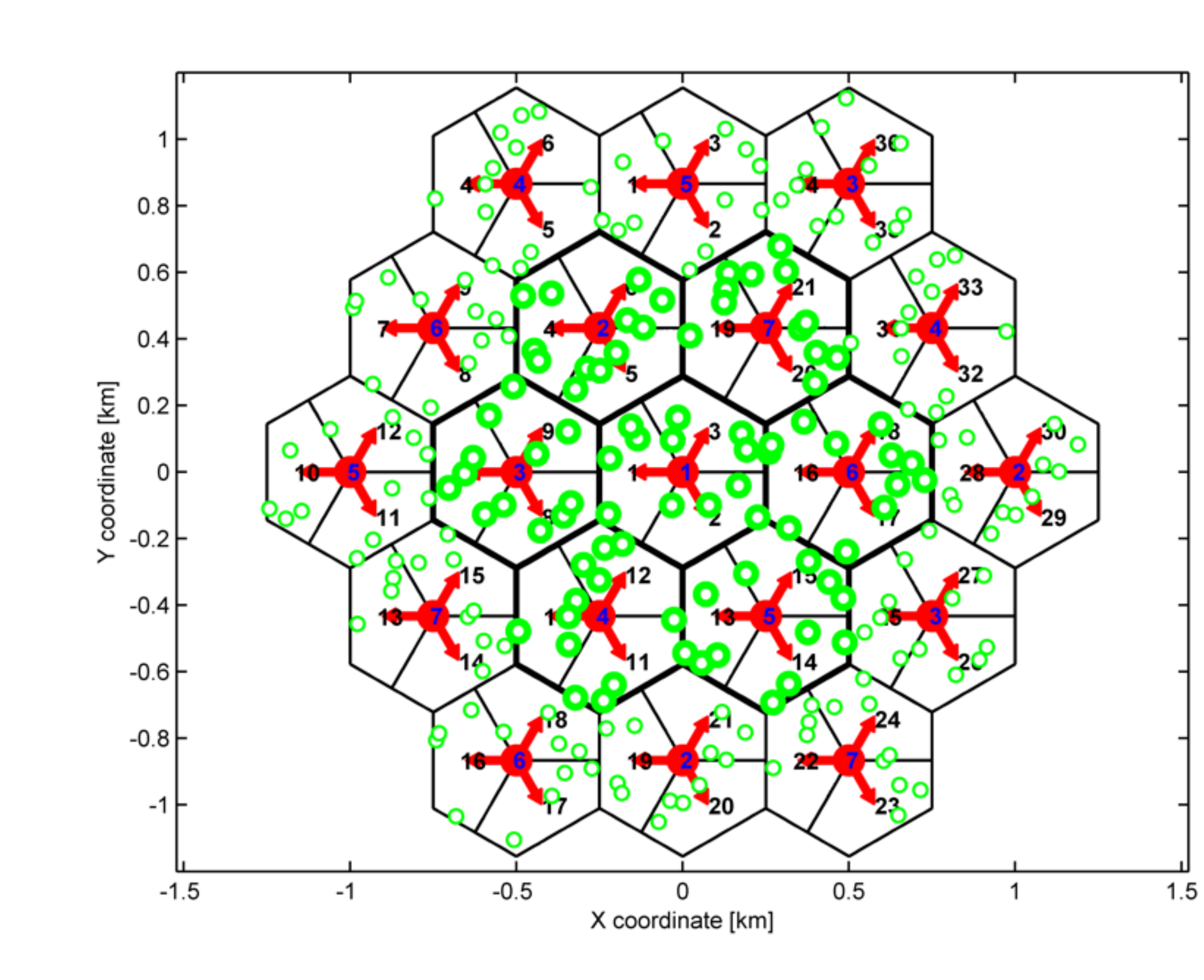}
\renewcommand{\figurename}{Fig.}
\caption{\label{fig:3gpp_mod_all_cells}Illustration of a typical 4G LTE SCN scenario.}
\vspace{-0.4cm}
\end{figure}

Megacities like New York are characterised by macrocell deployments with a 500$\,$m inter-site distance or less,
where each macrocell site is usually divided into 3$\,$macrocell sectors with directional antennas,
and each macrocell sector hosts a given number of small cell BSs.
Based on the above description,
and assuming today's 500$\,$m inter-site distance together with 4 small cells per macrocell sector~\cite{TR36.828},
we provide an illustration of a typical 4G LTE SCN scenario in Fig.~\ref{fig:3gpp_mod_all_cells},
which has
a small cell density of around $50\,\textrm{cells/km}{}^{2}$.
In this figure,
\emph{i)} each macrocell site is illustrated with thick black lines and is equally divided into 3$\,$macrocell sectors with their directional antennas shown as red arrows,
\emph{ii)} each small cell is illustrated with thick green circles, and
\emph{iii)} a wrap-around approach,
depicted as a third macrocell tier,
is considered to avoid the underestimation of interference in the outer-rim region~\cite{TR36.828}.

\begin{table}
\caption{\label{tab:key_param}Parameter values recommended by the 3GPP~\cite{TR36.828}}
\centering{}{\small{}}%
\scalebox{0.9}{
\begin{tabular}{|l|l|}
\hline
{\small{}Parameter definition} & {\small{}Parameter values}\tabularnewline
\hline
\hline
{\small{}The BS Tx power} & {\small{}24\,dBm (on a 10\,MHz bandwidth)}\tabularnewline
\hline
{\small{}The maximum UE Tx power} & {\small{}23\,dBm (on a 10\,MHz bandwidth)}\tabularnewline
\hline
{\small{}The AWGN power at BSs} & {\small{}-91\,dBm (on a 10\,MHz bandwidth)}\tabularnewline
\hline
{\small{}The AWGN power at UEs} & {\small{}-95\,dBm (on a 10\,MHz bandwidth)}\tabularnewline
\hline
{\small{}The BS-to-UE LoS path loss} & {\small{}$103.8+20.9\log_{10}r$ [dB] ($r$ in km)}\tabularnewline
\hline
{\small{}The BS-to-UE NLoS path loss\hspace{-0.1cm}} & {\small{}$145.4+37.5\log_{10}r$ [dB] ($r$ in km)}\tabularnewline
\hline
{\small{}The BS-to-BS LoS path loss} & {\small{}$101.9+40\log_{10}r$ [dB] ($r$ in km)}\tabularnewline
\hline
{\small{}The BS-to-BS NLoS path loss\hspace{-0.1cm}} & {\small{}$169.36+40\log_{10}r$ [dB] ($r$ in km)}\tabularnewline
\hline
{\small{}The UE-to-UE LoS path loss} & {\small{}$98.45+20\log_{10}r$ [dB] ($r$ in km)}\tabularnewline
\hline
{\small{}The UE-to-UE NLoS path loss\hspace{-0.1cm}} & {\small{}$175.78+40\log_{10}r$ [dB] ($r$ in km)}\tabularnewline
\hline
{\small{}The BS-to-UE LoS probability\hspace{-0.1cm}} & {\small{}\hspace{-0.1cm}\hspace{-0.1cm}$\begin{cases}
\hspace{-0.2cm}\begin{array}{l}
1\hspace{-0.1cm}-\hspace{-0.1cm}5\exp\left(\frac{-0.156}{r}\right)\hspace{-0.1cm},\hspace{-0.1cm}\\
5\exp\left(\frac{-r}{0.03}\right),
\end{array}\hspace{-0.3cm}\hspace{-0.1cm}\hspace{-0.1cm} & \begin{array}{l}
0<r\leq0.0677\,\textrm{km}\hspace{-0.1cm}\hspace{-0.1cm}\hspace{-0.1cm}\\
r>0.0677\,\textrm{km}
\end{array}\end{cases}$}\tabularnewline
\hline
{\small{}The BS-to-BS LoS probability\hspace{-0.1cm}} & {\small{}\hspace{-0.1cm}\hspace{-0.1cm}$\begin{cases}
\hspace{-0.2cm}\begin{array}{l}
1\hspace{-0.1cm}-\hspace{-0.1cm}5\exp\left(\frac{-0.156}{r}\right)\hspace{-0.1cm},\hspace{-0.1cm}\\
5\exp\left(\frac{-r}{0.03}\right),
\end{array}\hspace{-0.3cm}\hspace{-0.1cm}\hspace{-0.1cm} & \begin{array}{l}
0<r\leq0.0677\,\textrm{km}\hspace{-0.1cm}\hspace{-0.1cm}\hspace{-0.1cm}\\
r>0.0677\,\textrm{km}
\end{array}\end{cases}$}\tabularnewline
\hline
{\small{}The UE-to-UE LoS probability\hspace{-0.1cm}} & {\small{}\hspace{-0.1cm}\hspace{-0.1cm}$\begin{cases}
\hspace{-0.2cm}\begin{array}{l}
1,\\
0,
\end{array}\hspace{-0.3cm}\hspace{-0.1cm}\hspace{-0.1cm} & \begin{array}{l}
0<r\leq0.05\thinspace\textrm{km}\\
r>0.05\thinspace\textrm{km}
\end{array}\end{cases}$}\tabularnewline
\hline
\end{tabular}{\small \par}}
\vspace{-0.6cm}
\end{table}

In this paper, we consider a similar system model with macrocell sites deployed in a hexagonal grid to guide the random deployment of small cells with a density  of  $\lambda\,\textrm{BSs/km}^{2}$.
In particular,
we study three different small cell BS densities:
\begin{itemize}
  \item \emph{i)}
  {A typical 4G LTE network with 4 small cells per macrocell}, resulting in a $\lambda$ around $50\,\textrm{BSs/km}^{2}$;
  \item \emph{ii)}
  {A typical 5G dense SCN with 20 small cells per macrocell} (5-fold increase over the 4G SCN), resulting in a $\lambda$ around $250\,\textrm{BSs/km}^{2}$; and
  \item \emph{iii)}
  {A typical UDN with 200 small cells per macrocell} (50-fold increase over the 4G SCN), resulting in a $\lambda$ around $2500\,\textrm{BSs/km}^{2}$.
\end{itemize}

Moreover, we assume that active UEs are uniformly distributed in the considered network scenario with a density of $\rho\,\textrm{UEs/km}^{2}$.
Here, we only consider active UEs in the network because non-active UEs do not trigger data transmission,
and they are thus ignored in our study.
According to~\cite{Tutor_smallcell},
a typical density of the active UEs in 5G is around $\rho=300\thinspace\textrm{UEs/km}^{2}$.%

We also consider practical small cell BSs,
which implement idle mode capabilities (IMCs),
and thus can mute its wireless transmissions,
if there is no UE connected to it.
This reduces unnecessary inter-cell interference and energy consumption~\cite{dynOnOff_Huang2012}.
The set of active BSs should be determined by a user association strategy (UAS).
In this paper,
we assume a practical UAS as in~\cite{Tutor_smallcell},
where each UE is connected to the BS having
the minimum path loss,
which is equivalent to a maximum received signal strength policy when all cells in the scenario have the same transmit power.

As recommended in~\cite{TR36.828},
we consider practical LoS and non-line-of-sight (NLoS) transmissions,
and treat them as probabilistic events.
Both the LoS/NLoS path loss functions and the LoS probability function are defined based on a three-dimensional (3D) distance $r$,
i.e., $r=\sqrt{d^{2}+L^{2}}$,
where $d$ and $L$ denote the horizontal distance and the antenna difference between a transmitter and a receiver, respectively.
Note that the LoS probability decreases as $r$ increases~\cite{TR36.828}.

Finally, we assume that each BS and UE is equipped with an isotropic antenna,
and that the multi-path fading between a transmitter and a receiver is modelled as an independent and identically distributed (i.i.d.) Rayleigh random variable.
Additive white Gaussian noises (AWGN) are considered at the receiver side.
In this paper,
we adopt the power settings and the probabilistic LoS/NLoS path loss functions recommended by the 3GPP~\cite{TR36.828},
which are summarised in Table~\ref{tab:key_param}.

\begin{figure*}

\center

\subfloat[\label{fig:Hmap_BS_pos}Random BS deployments, shown as dots, for network visualisation.
As discussed in Section~\ref{sec:System-Model},
the BS densities of a typical 4G LTE network,
a typical 5G dense SCN,
and a typical UDN are around $50, 250, 2500\,\textrm{BSs/km}^{2}$,
respectively.]{\hspace{-1.6cm}\includegraphics[width=15.5cm]{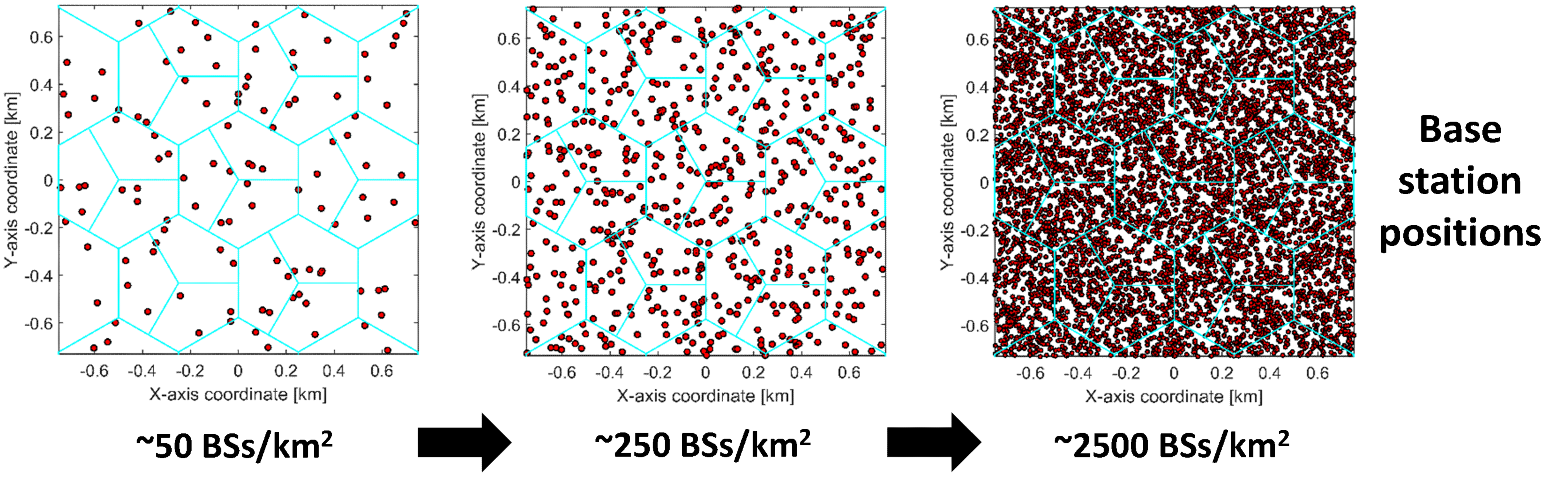}\vspace{-0.1cm}
}

\subfloat[\label{fig:Hmap_Jeff}Our understanding on the performance trend of UDNs before 2015 with a single-slope path loss function (NLoS only).
The sizes of the reddish areas (high SINR coverage probabilities) and the blueish areas (low SINR coverage probabilities) are approximately the same as the network densifies, indicating an invariant SINR performance in UDNs.]{\includegraphics[width=16cm]{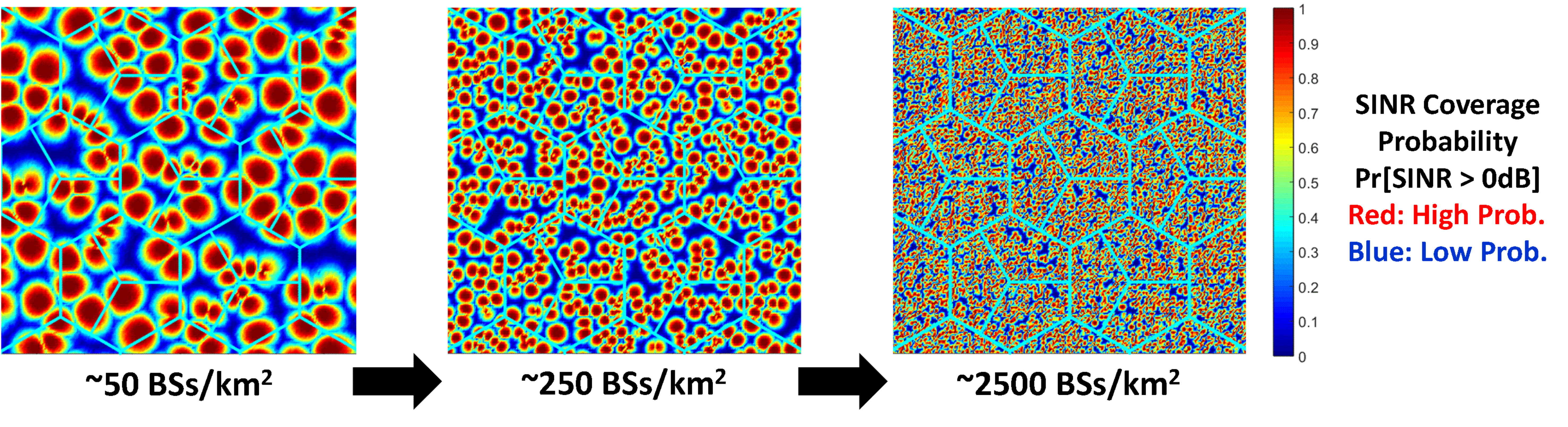}\vspace{-0.1cm}
}

\subfloat[\label{fig:Hmap_anteH}The performance trend of UDNs due to the non-zero BS-to-UE antenna height difference of $8.5\,\textrm{m}$~\cite{TR36.828}.
Compared with Fig.~\ref{fig:Hmap_Jeff}, the SINR heat map becomes blueish as $\lambda$ increases to around $2500\,\textrm{BSs/km}^{2}$,
showing a significant performance degradation because of the cap on the received signal power strength caused by the non-zero BS-to-UE antenna height difference in UDNs.
Such issue can be fixed by lowering the BS antenna height to the human height around $1.5\,\textrm{m}$.]
{\includegraphics[width=16cm]{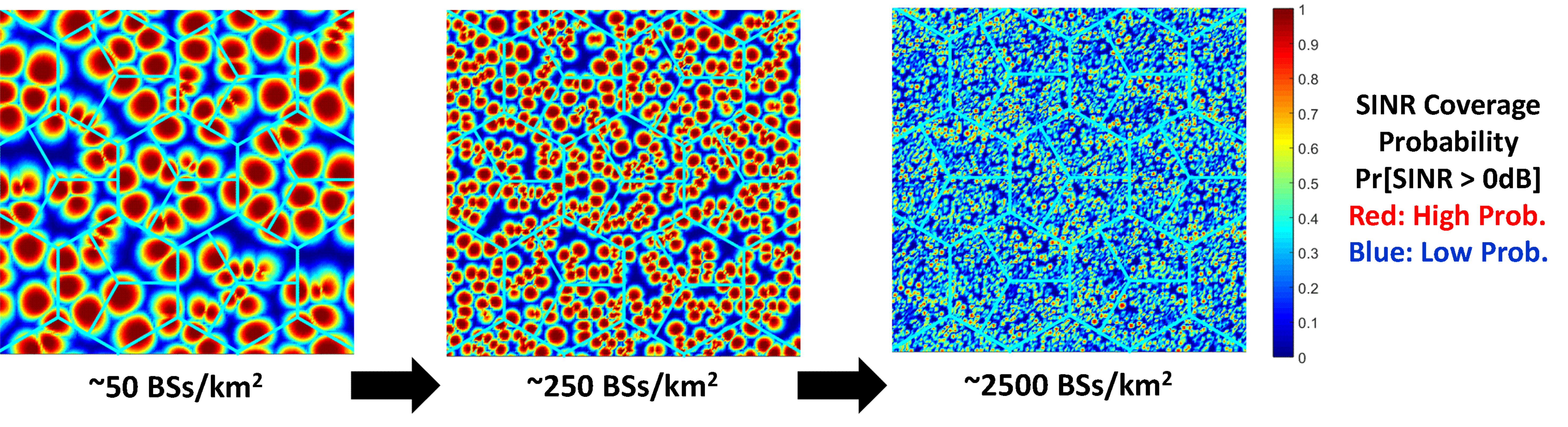}\vspace{-0.1cm}
}

\subfloat[\label{fig:Hmap_LoS}The performance trend of UDNs due to the high probabilities of LoS transmissions
(with the 3GPP path loss model incorporating both LoS and NLoS transmissions~\cite{TR36.828}).
Compared with Fig.~\ref{fig:Hmap_Jeff},
the SINR heat map becomes more and more blueish as $\lambda$ increases,
showing a significant performance degradation because of the transition
of a large number of interfering paths from NLoS to LoS in dense networks and UDNs.]
{\includegraphics[width=16cm]{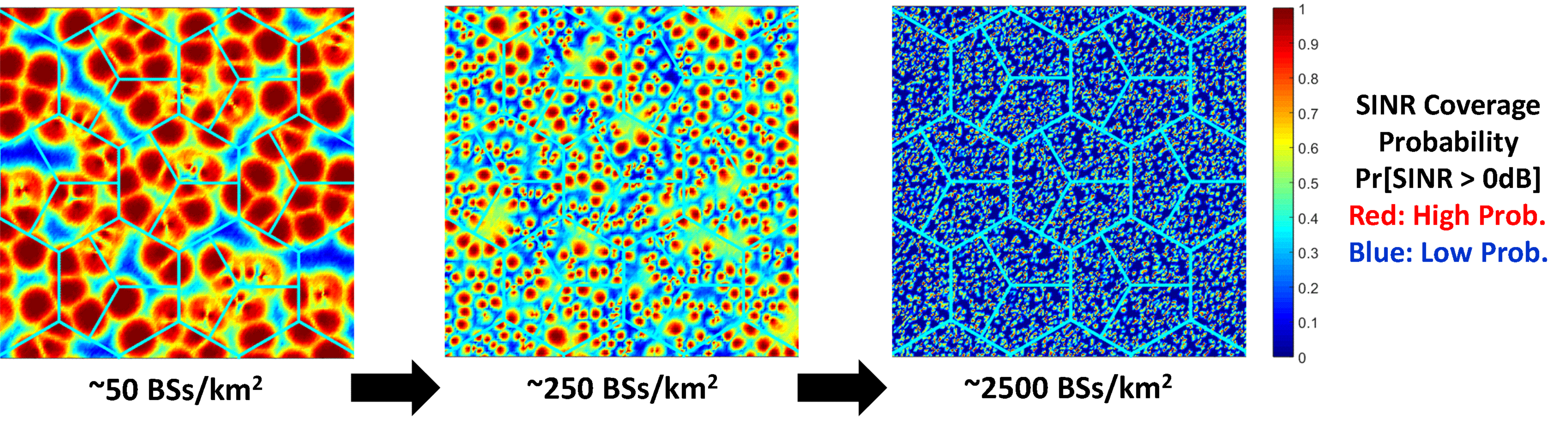}\vspace{-0.1cm}
}

\renewcommand{\figurename}{Fig.}\caption{\label{fig:PI_shorter_distance}
SINR heat maps considering a much shorter distance between a UE and its serving BS in UDNs.}

\vspace{-0.1cm}
\end{figure*}

\section{Network Performance Visualisation\label{sec:Study-Methodology}}

In this article,
instead of showing complex mathematical charts,
we propose an intuitive network performance visualisation technique to \emph{simulate and illustrate} the performance of UDNs and open our results to a wider audience.

First,
we randomly generate several network deployments following the 3GPP approach of the hexagonal grid addressed in Section~\ref{sec:System-Model}.
As discussed in Section~\ref{sec:System-Model},
the BS densities of a typical 4G LTE network,
a typical 5G dense SCN,
and a typical UDN are around $50, 250, 2500\,\textrm{BSs/km}^{2}$,
respectively.
In Fig.~\ref{fig:Hmap_BS_pos},
we plot the BS positions used in our network performance visualisation,
which are represented by dots and uniformly distributed over an area of $1.5\,\textrm{km}\times1.5\,\textrm{km}$.
Note that the hexagonal grid in Fig.~\ref{fig:Hmap_BS_pos} is of the same size as that in Fig.~\ref{fig:3gpp_mod_all_cells}.
Such 3GPP hexagonal grid is used here to guide the random deployment of small cell BSs in the network.

Second, based on the explicit modelling of the signal, the interference the noise powers addressed in Section~\ref{sec:System-Model},
we place a probe UE at a random position in the network,
and then measure its signal-to-interference-plus-noise ratio (SINR) coverage probability via \emph{10,000} simulation experiments. 
Such SINR coverage probability is defined as the probability that the probe UE's SINR is larger than a threshold $\gamma$, i.e., $\textrm{Pr}\left[\mathrm{SINR}>\gamma\right]$~\cite{Jeff2011}.
We repeat this probing process until the whole network is scanned with an adequate geographical resolution.
Moreover,
we set the SINR threshold $\gamma$ to 1 (i.e., 0\,dB),
which reflects the minimum working SINR threshold for practical receivers.

Third,
since the SINR coverage probability ranges between 0 and 1,
we show the network performance using heat maps,
with red and blue indicating a high SINR coverage probability and a low one, respectively.
With such SINR heat maps,
the significant performance impacts of the fundamental characteristics of UDNs can be translated into striking visual plots,
showing performance transition as the network densifies.

It should be noted that a large body of theoretical work has used the stochastic geometry theory to study network performance in terms of average coverage probability~\cite{Jeff2011,related_work_Jeff,Ding2016GC_ASECrash}.
Such performance metric averages the SINR coverage probability in the spatial domain and gives a single performance measure on a \emph{macroscopic} level.
In our network performance visualisation,
the average color of a heat map generated for a given scenario represents such average coverage probabilities.

Moreover,
the proposed network visualisation technique provides \emph{microscopic} information on network performance,
such as the \emph{variance} of the SINR coverage probability,
the locations of coverage holes, etc.,
which is useful in academic studies and for business proposals with clients.
For example,
the proposed technique is able to provide answers visually when an operator wants to know
the performance impact of adding several specific BSs to its current network.



\section{Two Fundamental Aspects of UDNs Related to BS-to-UE Distance\label{sec:two-fundamental-aspects-related-to-distance}}

Before 2015,
the common understanding on UDNs was that
the density of BSs would not affect the SINR coverage probability performance in interference-limited fully-loaded wireless networks,
and thus the network capacity in $\textrm{bps/Hz/km}^{2}$ would scale linearly with the network densification due to the linear increase of the spectrum reuse~\cite{Jeff2011}.


Fig.~\ref{fig:Hmap_Jeff} exemplifies such conclusion with a path loss exponent of 3.75,
showing the performance trend of three networks with different BS densities according to~\cite{Jeff2011}.
Visually speaking,
the sizes of the reddish areas (high SINR coverage probabilities) and the blueish areas (low SINR coverage probabilities) are approximately the same as the network densifies, indicating an invariant SINR performance with the BS density.

However, it is important to note that this conclusion was obtained with considerable simplifications on the propagation environment and BS/UE deployment,
such as a single-slope path loss model,
which should be placed under scrutiny when evaluating UDNs,
since they are fundamentally different from sparse ones in various aspects.
In this section, we study two fundamental aspects of UDNs,
both resulting from the much shorter distance between a BS and its served UEs.

\subsection{Capped Carrier Signal Power\label{subsec:A1_antH}}

In the performance analysis of the conventional sparse or dense cellular networks,
the antenna height difference between BSs and UEs is denoted by $L$,
and is usually ignored due to the dominance of the horizontal distance.
However, with a much shorter distance between a BS and its served UEs in an UDN,
such antenna height difference becomes non-negligible.
More specifically, the existence of a non-zero antenna height difference between BSs and UEs gives rise to a \emph{non-zero cap on the minimum distance} between a BS and its served UEs,
and thus \emph{a cap on the received signal power strength}.
Thus, and although each inter-cell interference power strength is subject to the same cap,
the aggregate inter-cell interference power will overwhelm the signal power in an UDN due to the sheer number of strong interferers.

To visualise this fundamental aspect,
in Fig.~\ref{fig:Hmap_anteH}
we plot the SINR heat map of SCNs with the following assumption:
\begin{itemize}
  \item the 3GPP antenna configuration using $L=8.5\,\textrm{m}$ as recommended in~\cite{TR36.828}. Note that in this case, the BS antenna height and the UE antenna height are assumed to be 10$\,$m and 1.5$\,$m, respectively.
\end{itemize}
\textbf{Obversations:}
From Fig.~\ref{fig:Hmap_anteH},
we can observe that:
\begin{itemize}
\item
The non-zero BS-to-UE antenna height difference has a negligible performance impact when the BS density is not ultra-dense
(i.e., when $\lambda$ is respectively around $50\,\textrm{BSs/km}^{2}$ and $250\,\textrm{BSs/km}^{2}$),
since the SINR maps are basically the same as those in Fig.~\ref{fig:Hmap_Jeff}.
\item
Compared with Fig.~\ref{fig:Hmap_Jeff},
the SINR heat map becomes much more blueish for the considered the UDN (i.e., when $\lambda$ is around $2500\,\textrm{BSs/km}^{2}$),
indicating a significant \emph{performance degradation}.
This is due to the cap on the received signal power strength caused by the non-zero BS-to-UE antenna height difference in UDNs.
\end{itemize}
\textbf{Concluding Remark:}
Considering the significant performance impact of the BS-to-UE antenna height difference,
we suggest to deploy small cell BSs in UDNs as closer as possible to human height.
However, this requires a revolutionised approach to BS architecture and deployment to avoid tampering, vandalism,
and other unwanted effects~\cite{Ding2016GC_ASECrash}.

In the following sections,
we will only consider a futuristic scenario with a zero antenna height difference between BSs and UEs,
i.e., $L=0\,\textrm{m}$,
so that the performance of UDNs is not unnecessarily penalised by this aspect as shown in Fig.~\ref{fig:Hmap_anteH}.

\begin{figure*}
\center

\subfloat[\label{fig:Hmap_IMC}The performance trend of UDNs due to the oversupply of BSs with IMC compared with UEs (the UE density is $\rho=300\thinspace\textrm{UEs/km}^{2}$).
Compared with Fig.~\ref{fig:Hmap_LoS},
the SINR heat map becomes very much reddish when $\lambda$ increases to around $2500\,\textrm{BSs/km}^{2}$,
showing a significant performance improvement in UDNs thanks to the BS IMC.
Such performance improvement is not obvious when $\lambda$ is around $50\,\textrm{BSs/km}^{2}$ since almost all BSs are active,
with or without the BS IMC.]
{\includegraphics[width=16cm]{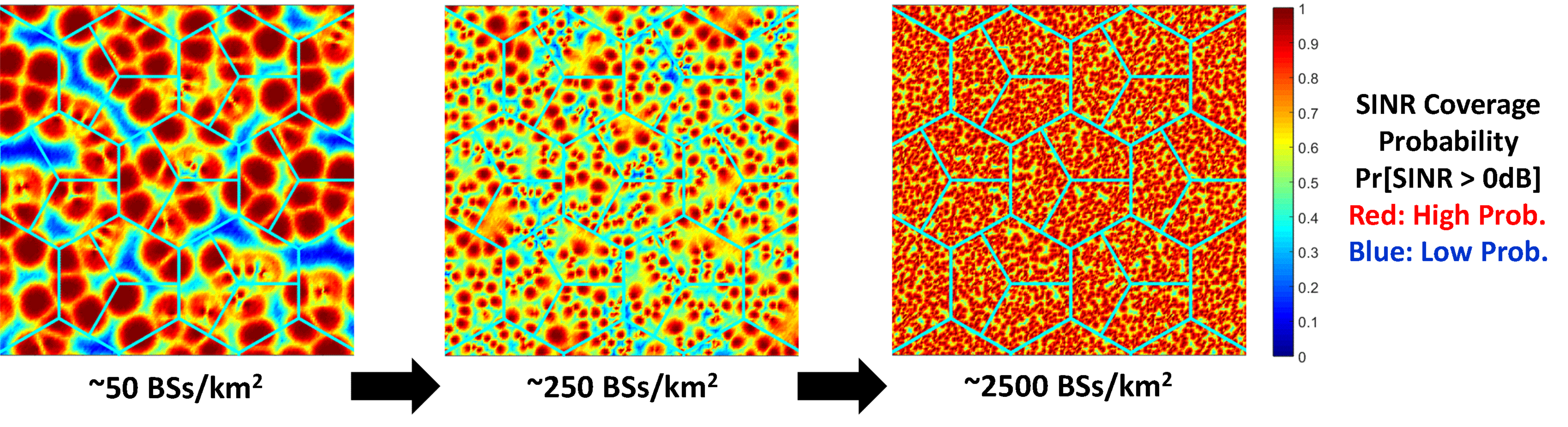}\vspace{-0.1cm}
}

\subfloat[\label{fig:Hmap_PF_DL}The performance trend of UDNs due to the PF scheduler working on a small number of UEs per active BS (the UE density is $\rho=300\thinspace\textrm{UEs/km}^{2}$).
Compared with Fig.~\ref{fig:Hmap_IMC},
the SINR heat map becomes reddish when $\lambda$ is lower than $250\,\textrm{BSs/km}^{2}$,
showing a performance improvement owing to a multi-UE diversity,
i.e., each BS can select a UE from a plurality of UEs to serve with a good channel quality.
Such performance gain is negligible when $\lambda$ increases to around $2500\,\textrm{BSs/km}^{2}$ due to the diminishing multi-UE diversity in UDNs.]
{\includegraphics[width=16cm]{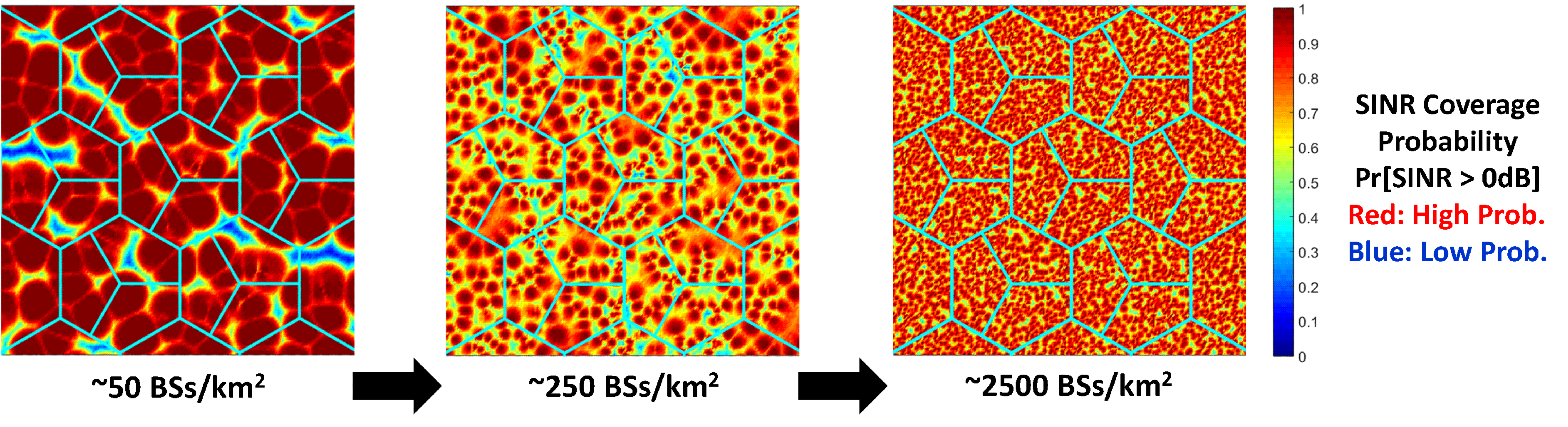}\vspace{-0.1cm}
}

\renewcommand{\figurename}{Fig.}\caption{\label{fig:PI_large_BS2UE_ratio}SINR heat maps considering a much larger ratio of the BS number over the UE number in UDNs.}

\vspace{-0.1cm}
\end{figure*}

\subsection{Stronger Interference Power\label{subsec:A2_LoS}}

A much shorter distance between a BS and its served UEs in an UDN implies high probabilities of line-of-sight (LoS) transmissions.
Generally speaking, LoS transmissions are more helpful to enhance the received signal strength than non-line-of-sight (NLoS) transmissions.
However, after a certain level of densification,
not only the signal power,
but also the inter-cell interference power will significantly grow due to the emergence of LoS paths.
Thus, the transition of a large number of interfering paths from NLoS to LoS will overwhelm the signal power in an UDN~\cite{related_work_Jeff}.

To visualise this fundamental aspect,
in Fig.~\ref{fig:Hmap_LoS}
we plot the SINR heat map of SCNs with the following assumptions:
\begin{itemize}
  \item the antenna configuration recommended in Subsection~\ref{subsec:A1_antH} with $L=0\,\textrm{m}$, and
  \item the 3GPP probabilistic LoS/NLoS path loss model exhibited in Table~\ref{tab:key_param}.
\end{itemize}
\textbf{Obversations:}
From Fig.~\ref{fig:Hmap_LoS},
we can see that:
\begin{itemize}
\item
Compared with Fig.~\ref{fig:Hmap_Jeff},
the SINR heat map becomes more reddish for the 4G SCN (i.e., when $\lambda$ is around $50\,\textrm{BSs/km}^{2}$),
because the received signal strength is enhanced by LoS transmissions,
while most inter-cell interference remain as NLoS.
\item
Compared with Fig.~\ref{fig:Hmap_Jeff},
the SINR heat map becomes more blueish when $\lambda$ is equal to or larger than $250\,\textrm{BSs/km}^{2}$,
thus showing a significant \emph{performance degradation} because of the transition of a large number of interfering paths from NLoS to LoS in UDNs.
\end{itemize}
\textbf{Concluding Remark:} Considering the significant performance impact of the transition of a large number of interfere paths from NLoS to LoS,
we propose that operators should carefully plan the deployment of small cell BSs from 4G to 5G,
and adopt potential solutions to mitigate the acceleration of the growth of the inter-cell interference,
which will be discussed later in this article.

In the following sections,
we will focus on
the 3GPP probabilistic LoS/NLoS path loss model exhibited in Table~\ref{tab:key_param}.
This propagation model is realistic and such NLoS to LoS transition cannot be avoided in the foreseeable future.

\section{Two Fundamental Aspects of UDNs Related to Small UE Number Per BS\label{sec:two-fundamental-aspects-related-to-UE-number}}

In this section, we study another two fundamental aspects of UDNs,
caused by the much larger ratio of the BS number over the UE number.

\subsection{More BSs than Active UEs\label{subsec:A3_IMC}}

In Section~\ref{sec:System-Model},
it was discussed how, with a surplus of BSs,
a BS can mute its transmissions when there is no active UE connected to it,
which is referred as the BS idle mode capability (IMC)~\cite{Tutor_smallcell}.
As a result, the active UE SINR performance benefits from
\emph{(i)} a BS diversity gain,
i.e., each UE has a plurality of BSs to select its server from, and
\emph{(ii)} a tight control of unnecessary inter-cell interference,
as idle BSs are put to sleep thanks to the IMC and do not generate any interference.
The surplus of BSs together with the IMC can be seen as a powerful tool that can help to mitigate the fast-growing interference problem
presented in Subsection~\ref{subsec:A2_LoS}.

To visualise this fundamental aspect,
in Fig.~\ref{fig:Hmap_IMC}
we plot the SINR heat map of SCNs with the following assumptions:
\begin{itemize}
  \item the antenna configuration with $L=0\,\textrm{m}$ recommended in Subsection~\ref{subsec:A1_antH},
  \item the 3GPP probabilistic LoS/NLoS path loss model recommended in Subsection~\ref{subsec:A2_LoS}, and
  \item a finite number of UEs ($\rho=300\thinspace\textrm{UEs/km}^{2}$,
  a typical density of the active UEs in 5G)~\cite{Tutor_smallcell}.
\end{itemize}
\textbf{Obversations:}
From Fig.~\ref{fig:Hmap_IMC},
we can see that:
\begin{itemize}
\item
Compared with Fig.~\ref{fig:Hmap_LoS},
the SINR heat map changes little for the 4G SCN (i.e., when $\lambda$ is around $50\,\textrm{BSs/km}^{2}$),
because almost all BSs are active with or without the IMC,
due to the relatively small BS density compared with the UE density of $\rho=300\thinspace\textrm{UEs/km}^{2}$.
\item
Compared with Fig.~\ref{fig:Hmap_LoS},
the SINR heat map becomes more reddish when $\lambda$ is equal to or larger than $250\,\textrm{BSs/km}^{2}$,
thus showing a significant \emph{performance improvement} in UDNs thanks to the BS IMC.
More specifically,
the signal power continues increasing with the network densification,
while the interference power is controlled because not all BSs are turned on.
\end{itemize}
\textbf{Concluding Remark:}
Considering the significant performance benefit of the surplus of BS together with the IMC,
we suggest that small cell BSs in UDNs should be equipped with efficient IMCs that are able to put to sleep a given number of BSs to realise a desired UE quality of experience (QoE).

In the following sections,
we will consider the BS IMC because it is essential to the successful operation of UDNs.

\subsection{Less active UEs per active BS\label{subsec:A4_scheduling}}

With a much larger number of BSs than UEs in UDNs,
the number of active UEs in each BS also  becomes smaller,
and approaches the one-active-UE-per-BS limit when the BS density is very large~\cite{Tutor_smallcell}.
Such phenomenon incurs in a diminishing multi-UE diversity gain,
which refers to the gain that each BS can select a UE from a plurality of UEs to serve with a good channel quality.
When the one-active-UE-per-BS limit is reached,
then one BS is stuck with one UE and cannot perform any UE selection in its scheduler.
In practice,
two commonly used BS schedulers are
\emph{(i)} the round-robin (RR) scheduler,
which selects UEs in turn regardless of their channel state, and
\emph{(ii)} the proportional fair (PF) scheduler that tends to select UEs with larger multi-path fading
while reinforcing a certain degree of fairness among UEs~\cite{Tutor_smallcell}.
However,
as the number of active UEs per active BS approaches one in UDNs,
there will be very little gain that can be achieved by the advanced PF scheduler compared with the simplistic RR scheduler~\cite{Tutor_smallcell}.

To visualise this fundamental aspect,
in Fig.~\ref{fig:Hmap_PF_DL}
we plot the SINR heat map of SCNs with the following assumptions:
\begin{itemize}
  \item the antenna configuration with $L=0\,\textrm{m}$ recommended in Subsection~\ref{subsec:A1_antH},
  \item the 3GPP probabilistic LoS/NLoS path loss model recommended in Subsection~\ref{subsec:A2_LoS},
  \item the BS IMC recommended in Subsection~\ref{subsec:A3_IMC}, and
  \item the PF scheduler. Note that previous results in Figs.~\ref{fig:PI_shorter_distance} and~\ref{fig:Hmap_IMC} assumed a RR scheduler.
\end{itemize}
\textbf{Obversations:}
From Fig.~\ref{fig:Hmap_PF_DL},
we can see that:
\begin{itemize}
\item
Compared with Fig.~\ref{fig:Hmap_IMC},
the SINR heat map becomes more reddish when the BS density is not ultra-dense,
i.e., $\lambda$ is equal to or less than $250\,\textrm{BSs/km}^{2}$,
showing a performance improvement owing to the multi-UE diversity,
i.e., each BS can select a UE from a plurality of UEs to serve with a good channel quality.
\item
Compared with Fig.~\ref{fig:Hmap_IMC},
the SINR heat map changes little when $\lambda$ is around $2500\,\textrm{BSs/km}^{2}$,
thus showing \emph{a negligible performance gain} since such UDN approaches the one-active-UE-per-BS limit.
\end{itemize}
\textbf{Concluding Remark:}
Considering the negligible gain of the more complex channel-dependent PF scheduling in UDNs,
when complexity is an issue,
it is recommended to adopt simpler scheduling mechanisms such as the RR scheduler to simplify the radio resource management (RRM).

In the following sections,
as we are not analysing complexity but performance,
we will still consider the PF scheduler.
It inflicts no harm to UDNs,
while it has been widely adopted in sparse networks to enhance the performance.

\section{One Fundamental Aspect of UDNs Related to Busty Traffic Demands\label{sec:one-fundamental-aspect-related-to-bursty-traffic}}

As a by-product of the small number of active UEs in each active BS,
the per-cell aggregate traffic demands in the downlink (DL) and the uplink (UL) become highly dynamic.
Consequently,
dynamic TDD emerges as a new technology for UDNs,
which serves as a transition link between the conventional network and the full duplex one~\cite{Shen2012dynTDD,Ding2016dynTDD,Suciu2016SDR}.
In dynamic TDD,
each BS can provide a tailored configuration of DL/UL subframe resources to match the DL/UL data requests~\cite{TR36.828}.
However, such flexibility comes at the expense of allowing inter-cell inter-link interference (ICILI),
i.e., DL transmissions of a cell may interfere with UL ones of neighbouring cells (DL-to-UL interference) and vice versa (UL-to-DL interference).
This performance degradation is particularly severe for the UL due to the relatively strong DL-to-UL interference,
which calls for the implementation of interference cancellation (IC) and UL power boosting (ULPB) techniques~\cite{Shen2012dynTDD,Ding2016dynTDD}.

In more detail,
the IC technique leverages the backhaul communications among BSs and the BS signal processing capability to mitigate the DL-to-UL interference coming from other BSs.
Moreover,
a UL power control algorithm~\cite{TR36.828} is currently adopted in the LTE networks and considered in our network performance visualization.
For a typical example,
such algorithm partially compensates each UE's path loss (in dB) by 80\,\%,
on top of a base power of -59dBm for a 10MHz bandwidth.
If such power exceeds the UE power limit,
then it will be clipped to the maximum UE transmission power of 23\,dBm shown in Table~\ref{tab:key_param}.
A straightforward ULPB scheme would be letting UEs transmit at their full powers to combat the DL-to-UL interference.

\begin{figure*}
\center

\subfloat[\label{fig:Hmap_dynTDD_DL}The performance trend of UDNs with dynamic TDD (DL part).
Compared with Fig.~\ref{fig:Hmap_PF_DL},
the DL SINR performance of dynamic TDD is shown to be robust to ICILI for all BS densities.]
{\includegraphics[width=16cm]{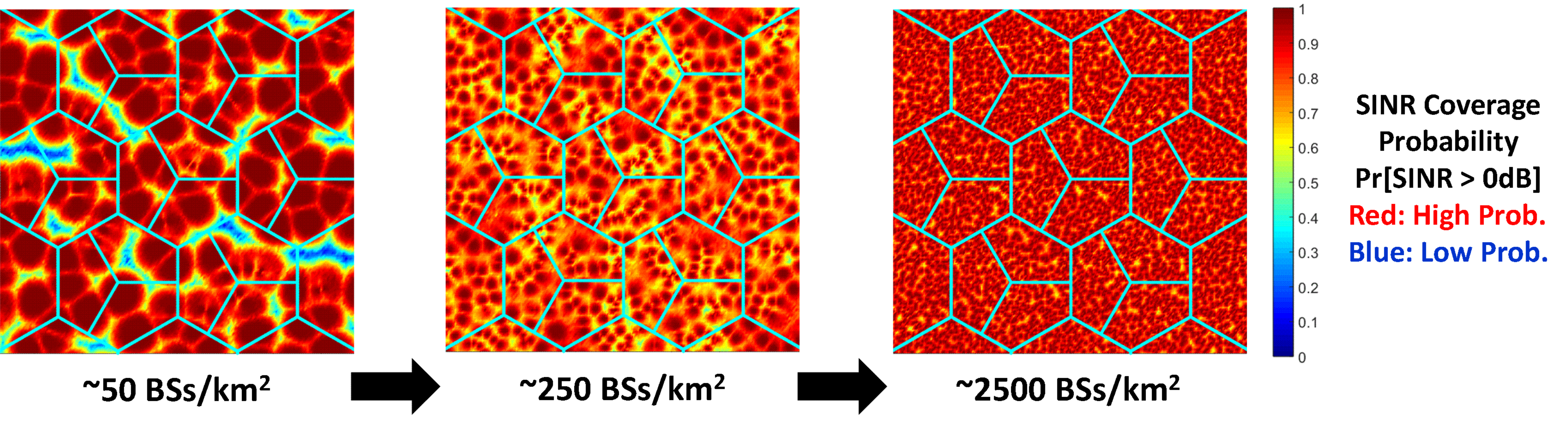}\vspace{-0.1cm}
}

\subfloat[\label{fig:Hmap_dynTDD_UL}The performance trend of UDNs with dynamic TDD (UL part).
The UL of dynamic TDD shows total outage for UDNs.]
{\includegraphics[width=16cm]{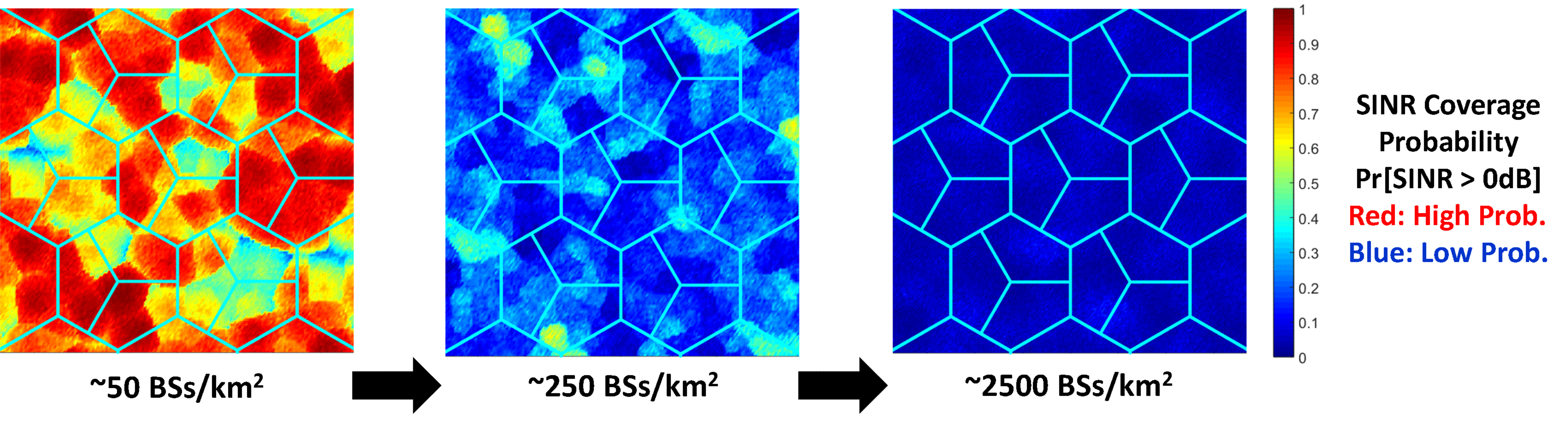}\vspace{-0.1cm}
}

\subfloat[\label{fig:Hmap_dynTDD_UL_IC3BS}The performance trend of UDNs with dynamic TDD (UL part with partial IC).
Compared with Fig.~\ref{fig:Hmap_dynTDD_UL},
the UL SINR performance is greatly improved for all BS densities,
thanks to the cancellation of the top 3 interfering signals from other BSs.]
{\includegraphics[width=16cm]{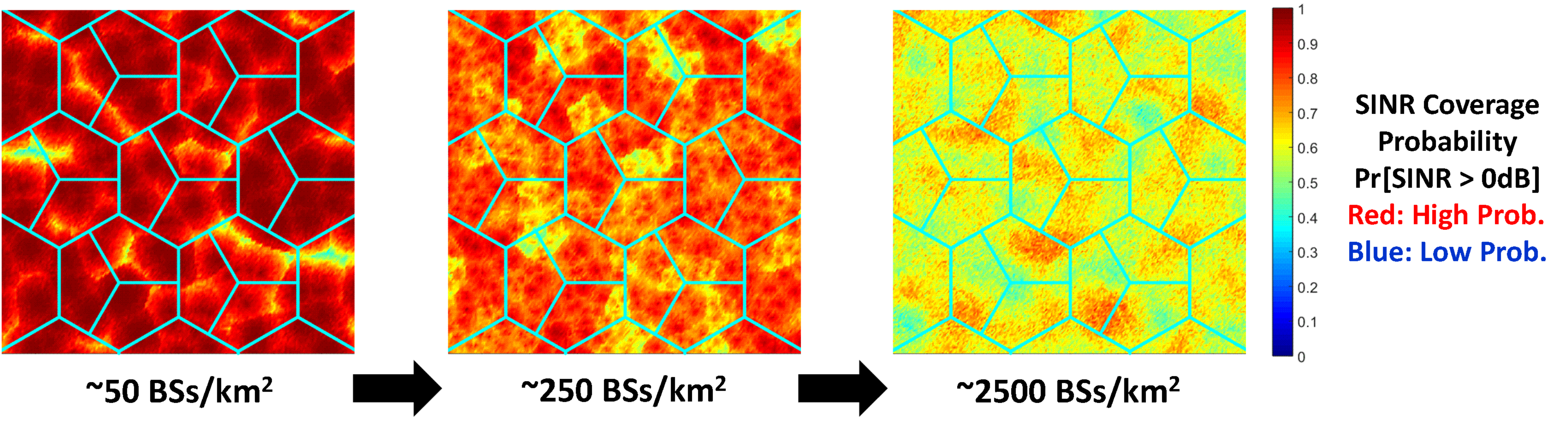}\vspace{-0.1cm}
}

\subfloat[\label{fig:Hmap_dynTDD_UL_IC3BS_ULPB}The performance trend of UDNs with dynamic TDD (UL part with partial IC + ULPB).
Compared with Fig.~\ref{fig:Hmap_dynTDD_UL_IC3BS},
the UL SINR performance is further improved,
due to the use of full power at UEs (23\,dBm).
Now the UL performance is comparable to the DL one shown in Fig.~\ref{fig:Hmap_dynTDD_DL}.]
{\includegraphics[width=16cm]{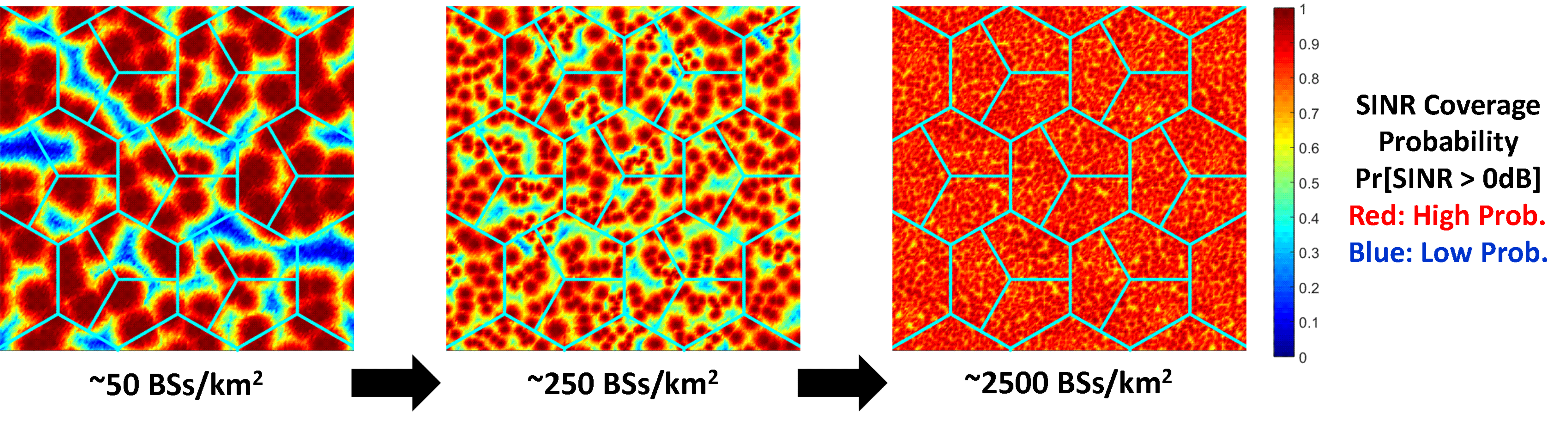}\vspace{-0.1cm}

}\renewcommand{\figurename}{Fig.}\caption{\label{fig:PI_bursty_traffic}SINR heat maps considering busty traffic demands in the DL and the UL.}

\vspace{-0.1cm}
\end{figure*}

To visualise this fundamental aspect,
in Fig.~\ref{fig:PI_bursty_traffic},
we plot the SINR heat map of SCNs with the following assumptions:
\begin{itemize}
  \item the antenna configuration with $L=0\,\textrm{m}$ recommended in Subsection~\ref{subsec:A1_antH},
  \item the 3GPP probabilistic LoS/NLoS path loss model recommended in Subsection~\ref{subsec:A2_LoS},
  \item the BS IMC recommended in Subsection~\ref{subsec:A3_IMC},
  \item the PF scheduler studied in Subsection~\ref{subsec:A4_scheduling}, and
  \item the LTE dynamic TDD with each UE randomly requesting DL data and UL one with 50\,\% and 50\,\%, respectively.
\end{itemize}
\textbf{Obversations:}
From Fig.~\ref{fig:PI_bursty_traffic},
we can see that:
\begin{itemize}
\item
In Fig.~\ref{fig:Hmap_dynTDD_DL},
the SINR heat map of the DL UDN (i.e., when $\lambda$ is around $2500\,\textrm{BSs/km}^{2}$) is slightly more reddish than Fig.~\ref{fig:Hmap_PF_DL},
because some DL interference is converted to UL interference,
which is generally weaker.
\item
In Fig.~\ref{fig:Hmap_dynTDD_UL},
the SINR heat map of the UL shows that the UL is particularly vulnerable to the DL-to-UL interference in UDNs,
especially for UDNs (i.e., when $\lambda$ is around $2500\,\textrm{BSs/km}^{2}$).
\item
A potential solution to mitigate the DL-to-UL interference is the partial IC technique.
Here, we assume that each BS can remove the top 3 interfering signals from other BSs.
In Fig.~\ref{fig:Hmap_dynTDD_UL_IC3BS},
the SINR heat map of the UL with partial IC shows considerable improvement than that in Fig.~\ref{fig:Hmap_dynTDD_UL},
especially for the 4G SCN and the 5G dense SCN (i.e., when $\lambda$ is equal to or less than $250\,\textrm{BSs/km}^{2}$).
\item
Note that no UL power boosting was considered in Figs.~\ref{fig:Hmap_dynTDD_UL} and~\ref{fig:Hmap_dynTDD_UL_IC3BS}.
Here,
we assume that each UE transmits at its full power,
i.e., 23\,dBm (see Table~\ref{tab:key_param})~\cite{TR36.828}.
In Fig.~\ref{fig:Hmap_dynTDD_UL_IC3BS_ULPB},
the SINR heat map of the UL with partial IC and ULPB is further improved when $\lambda$ is around $2500\,\textrm{BSs/km}^{2}$ compared with Fig.~\ref{fig:Hmap_dynTDD_UL_IC3BS}.
However,
such performance improvement is not obvious for the 4G SCN and the 5G dense SCN (i.e., when $\lambda$ is equal to or less than $250\,\textrm{BSs/km}^{2}$).
This is because the ULPB technique considerably increases the UL-to-UL inter-cell interference,
which makes cell edge UEs suffer.
\end{itemize}
\textbf{Concluding Remark:}
Considering the significant performance benefit of dynamic TDD,
it is recommended to enable dynamic TDD in UDNs.
However, special attention needs to be paid to mitigate the strong DL-to-UL interference.
The partial IC and the ULPB techniques were considered here.%

\section{Conclusion\label{sec:Conclusion}}

Using the proposed network performance visualisation technique,
we have investigated in a unified framework the five fundamental differences between current macrocell and sparse small cell networks and future UDNs.
From our results,
we conclude that in UDNs:
\begin{itemize}
\item
The small cell BS antenna height should be lowered closer to the UE antenna height to avoid the \emph{performance degradation}
due to the capped carrier signal power.
\item
The small cell BS density should be carefully considered to mitigate the \emph{performance degradation}
due to the NLoS to LoS interference
transition.
\item
The small cell BS IMC is a fundamental feature to unveil the full potential of UDNs
and allow the \emph{performance improvement}
due to the UE selection of a good serving BS and the control of unnecessary inter-cell interference.
\item
The RR scheduler could be used to simplify the resource management,
which is justified by its \emph{comparable performance} to the PF one in UDNs.
\item
The dynamic TDD technology could be used to harness the high dynamics of DL/UL traffic and provide the \emph{performance improvement} in the MAC layer.
However, an efficient interference management is needed in the PHY layer,
primarily to mitigate DL to UL interference.
\end{itemize}

As future work,
we propose the following:
\begin{itemize}
  \item The proposed network performance visualization technique can be extended for a 3D space using plane slicing.
  Such extension would be useful to illustrate network performance for scenarios with buildings or drones.
  \item As indicated in Section~\ref{sec:System-Model},
  we have considered uniform distributions of UEs and BSs in this article.
  It is of great interest to investigate a hotspot deployment with non-uniform distributions of UEs and BSs,
  which would change the distributions of the active BSs and the UL interfering UEs, etc.
  \item The BS IMC is very useful to reduce the energy consumption of UDNs.
  However,
  it is important to note that a BS in idle mode may still consume a non-negligible amount of energy,
  thus impacting the energy efficiency of small cell networks.
  To study the energy efficiency of realistic 5G networks,
  we can use the practical power model developed in the Green-Touch project~\cite{Tutor_smallcell}.

\end{itemize}
\bibliographystyle{IEEEtran}
\bibliography{Ming_library}

\noindent \vspace{-10cm}
\begin{IEEEbiographynophoto} {\textbf{Ming Ding}} is a Senior Research Scientist at Data61 (previously known as NICTA), CSIRO, Australia. He has authored over 60 papers in IEEE journals and conferences, all in recognized venues, and about 20 3GPP standardization contributions, as well as a Springer book "Multi-point Cooperative Communication Systems: Theory and Applications" (2013). As the first inventor, he holds 15 CN, 7 JP, 3 US, 2 KR patents and co-authored another 100+ patent applications on 4G/5G technologies.

\vspace{0.3cm}
\noindent
{\textbf{David L$\acute{\textrm{o}}$pez-P$\acute{\textrm{e}}$rez}} is currently a member of Technical Staff at Nokia Bell Laboratories. David has authored the book "Heterogeneous Cellular Networks: Theory, Simulation and Deployment" (Cambridge University Press, 2012), as well as over 100 book chapters, journal, and conference papers, all in recognized venues. He also holds over 36 patents applications. David received the Ph.D. Marie-Curie Fellowship in 2007 and the IEEE ComSoc Best Young Professional Industry Award in 2016.

\vspace{0.3cm}
\noindent
{\textbf{Holger Claussen}} is the Leader of Small Cells Research at Nokia Bell Labs. He is author of more than 70 publications and 100 filed patent applications. Moreover, he is Fellow of the World Technology Network, and member of the IET. He received the 2014 World Technology Award in the individual category Communications Technologies for innovative work of "the greatest likely long-term significance".

\vspace{0.3cm}
\noindent
{\textbf{Mohamed Ali Kaafar}} is the head of the Networks group in CSIRO Data61 and a senior principal research scientist. His main research interests include online privacy and privacy enhancing technologies, cyber-security and networks measurement and modelling.
\end{IEEEbiographynophoto}

\end{document}